%% file: MAlexander-CharmAtLHCb.tex
\documentclass{ws-ijmpcs}

\def\inputdir{.}
\input{\inputdir/lhcb-symbols-preamble}
\input{\inputdir/lhcb-symbols-def}

\input{\inputdir/analysis-symbols-def}

\input{\inputdir/general-latex.tex}

\def\rskpi{\decay{\Dz}{\Km\pip}}
\def\wskpi{\decay{\Dz}{\Kp\pim}}
\def\bothkpi{\decay{\Dz}{\Kpm\pimp}}
\def\dstdpi{\decay{\Dstarp}{\Dz\pip_s}}
\def\xpsq{\ensuremath{x^{\prime 2}}\xspace}
\def\yp{\ensuremath{y^{\prime}}\xspace}
\def\rd{\ensuremath{R_D}\xspace}
\def\mdstar{\ensuremath{m_{\Dstarp}}\xspace}
\def\md{\ensuremath{m_{\Dz}}\xspace}
\def\dtopimumu{\decay{\Dp_{(s)}}{\pip\mumu}}
\def\dtopimumuss{\decay{\Dp_{(s)}}{\pim\mup\mup}}
\def\dtopimumuboth{\decay{\Dp_{(s)}}{\pipm\mup\mmu^{\mp}}}
\begin{document}

\markboth{M. Alexander}
{Charm Physics at LHCb}

%
\catchline{}{}{}{}{}
%

\title{CHARM: MIXING, CP VIOLATION AND \\RARE DECAYS AT LHCB}

\author{MICHAEL ALEXANDER\footnote{232b Kelvin Building, University of Glasgow, Glasgow, UK.}}

\address{Physics \& Astronomy, University of Glasgow, \\
Glasgow, G12 8QQ,
United Kingdom\\
michael.alexander@glasgow.ac.uk\\
\mbox{}\\
On behalf of the LHCb collaboration}

\maketitle

\begin{history}
\received{$30^{th}$ November 2013}
\revised{$30^{th}$ November 2013}
\end{history}

\begin{abstract}
Recent results on mixing, CP violation and rare decays in charm physics from the \lhcb experiment are presented. Study of \dquote{wrong-sign} \wskpi decays provides the highest precision measurements to date of the mixing parameters \xpsq and \yp, and of CP violation in this decay mode. Direct and indirect CP violation in the \Dz system are probed to a sensitivity of around \tene{-3} using \decay{\Dz}{\KpKm} and \decay{\Dz}{\pipi} decays and found to be consistent with zero. Searches for the rare decays \dtopimumu, \dtopimumuss and \decay{\Dz}{\mumu} find no evidence of signal, but set the best limits on branching fractions to date. Thus, despite many excellent results in charm physics from \lhcb, no evidence for physics beyond the Standard Model is found.
\end{abstract}

\ccode{PACS numbers:13.25.Ft, 14.40.Lb}


\section{Introduction}
\label{sec:intro}
CP violation and the branching fractions of rare decays in charm physics can be significantly enhanced beyond Standard Model (SM) predictions by the presence of non-SM particles. Precision measurements of these parameters can thus provide evidence for, or strict limits on, new physics.

The \lhcb detector\cite{JINST_LHCb} at the LHC is specifically designed for precision measurements of CP violation in decays involving \c and \b quarks. The Vertex Locator provides fine tracking around the interaction point, achieving decay-time resolutions of around 50 \fs for \Dz mesons; the full tracking system yields momentum resolutions of $\sigma(\ptot)/\ptot \sim 0.5 \%$; and the two Ring Imaging Cherenkov detectors provide clean separation of pions and kaons. Additionally, the \ccbar production cross section in the collisions provided by the LHC is very large. With 1.1 \invfb integrated luminosity recorded at \sqrtseq{7 \tev} in 2011 and 2.1 \invfb at \sqrtseq{8 \tev} in 2012 this makes \lhcb an excellent source of high statistics datasets for the study of charm physics.


\section{Mixing and CP Violation in \bothkpi Decays}
\label{sec:WSKpi}

The \dquote{right-sign} (RS) decay of \rskpi\footnote{Charge conjugate states are implied throughout.} occurs predominantly through a Cabibbo-favored (CF) process in which no \Dz-\Dzb mixing occurs. The \dquote{wrong-sign} (WS) \wskpi decay occurs with roughly equal rate via a doubly-Cabibbo-suppressed (DCS) process and one in which the \Dz meson first mixes and then decays via the CF process. The ratio of the decay rate of WS decays to that of RS decays as a function of the decay time of the \Dz meson is thus sensitive to the mixing parameters \xpsq and \yp and is given by
\begin{equation}
  R(t) = \frac{N_{WS}(t)}{N_{RS}(t)} = \rd + \sqrt{\rd} \yp t + \frac{\xpsq + y^{\prime 2}}{4} t^2,
\end{equation}
where
\begin{equation*}
  \rd = \magsq{\frac{A_{DCS}}{A_{CF}}},\, x^{\prime} = x \cos(\delta) + y \sin(\delta),\, \yp = - x \sin(\delta) + y \cos(\delta),\,
\end{equation*}
\begin{equation}
  \delta = \mathrm{arg}\left(\frac{A_{DCS}}{A_{CF}}\right),\, x = \frac{\Delta m_{\Dz}}{\Gamma_{\Dz}},\, y = \frac{\Delta \Gamma_{\Dz}}{2\Gamma_{\Dz}},
  \label{eq:rt}
\end{equation}
$A_{DCS(CF)}$ is the amplitude of the DCS (CF) decay, $\Delta m_{\Dz}$ is the mass difference of the mass eigenstates of the \Dz system (defined as $\ket{\Dz_{H,L}} = p\ket{\Dz} \pm q \ket{\Dzb}$), $\Delta \Gamma_{\Dz}$ the decay width difference, and $\Gamma_{\Dz}$ the decay width of the \Dz meson. Any discrepancy in $R(t)$ between initial states of \Dz and \Dzb would indicate CP violation.

$R(t)$ is measured by firstly reconstructing and selecting \dstdpi candidates with \bothkpi\cite{lhcb:DMixing2013}. The charge of the $\pip_s$ track gives the flavor of the \Dz candidate at production. Backgrounds, predominantly from \Dz decays associated with a random $\pip_s$ track, are distinguished from signal by the distribution of the invariant mass of the \Dstarp candidates (\mdstar). An additional background arises from \Dz mesons produced in decays of \decay{\PB}{\Dz X}. These are strongly suppressed by a cut on the impact parameter (IP) \chisq of the \Dz candidate and a systematic uncertainty assigned to any remaining contribution. WS and RS candidates are divided into bins of \Dz decay time and the distribution of \mdstar fitted in each bin to obtain the yields. The ratio of WS to RS yields is then plotted against decay time to give $R(t)$ and fitted with Eq. \ref{eq:rt} to determine $\rd$, $\xpsq$ and $\yp$.

\begin{figure}
  \includegraphics[width=\textwidth]{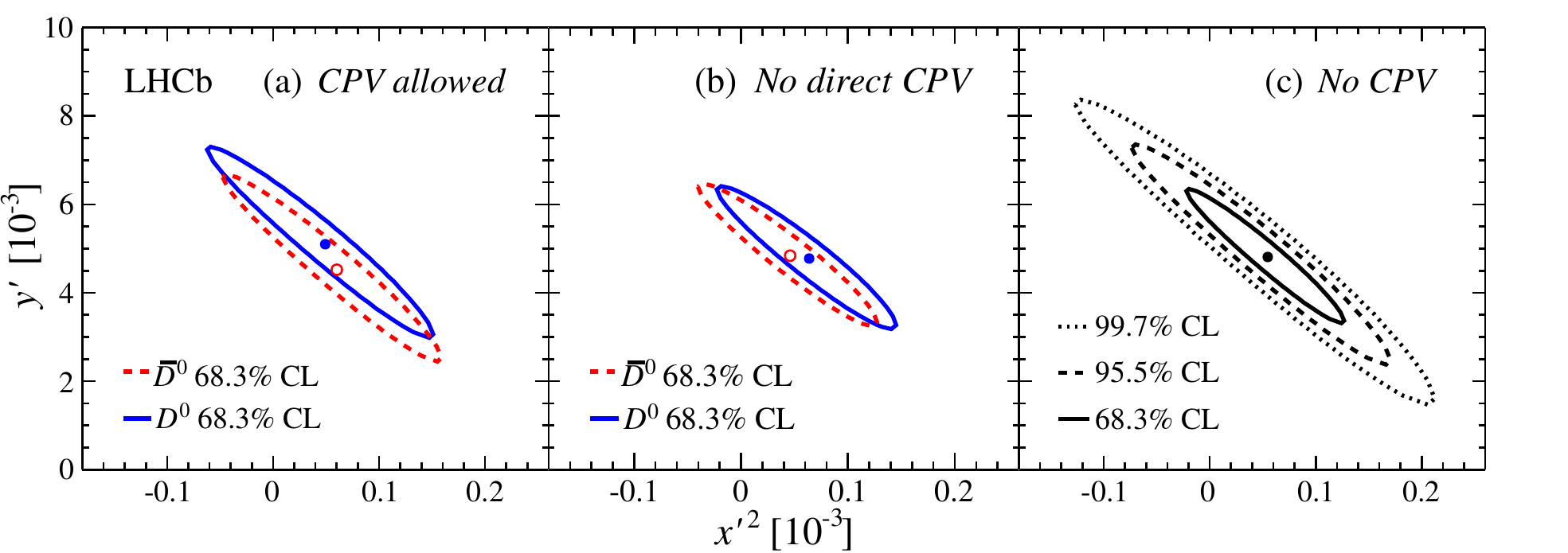}
  \caption[Results for \xpsq and \yp]{The results of fits for \xpsq and \yp in the case that (left) CP violation is allowed, (middle) only indirect CP violation is allowed, and (right) no CP violation is allowed.}
  \label{fig:WSKpiContours}
\end{figure}

The results of fits for \xpsq and \yp from the full 2011 and 2012 datasets are shown in Fig. \ref{fig:WSKpiContours}. The cases in which CP violation is allowed, only indirect CP violation is allowed, and no CP violation is allowed are all considered. The CP conservation case gives
\begin{equation*}
  \rd = \xtene{(3.568 \pm 0.066)}{-3},\, \xpsq = \xtene{(5.5 \pm 4.9)}{-5},\, \yp = \xtene{(4.8 \pm 1.0)}{-3},
\end{equation*}
which excludes the no mixing hypothesis at $> 10 \sigma$. Additionally, the cases for which CP violation is allowed find 
\begin{equation*}
  \frac{\rd(\Dz) - \rd(\Dzb)}{\rd(\Dz) + \rd(\Dzb)} = (-0.7 \pm 1.9) \%, 
\end{equation*}
and $0.75 < \magnitude{q/p} < 1.24$ at 68.3 \% confidence. These are the most precise single measurements of these parameters to date and show no evidence for CP violation.

\section{Indirect CP Violation in \decay{\Dz}{\hphm} Decays}
\label{sec:agamma}
The parameter \agamma, defined below, gives access primarily to indirect CP violation as
\begin{equation}
  \agamma = \frac{\taueff{\Dzb}{\hphm} - \taueff{\Dz}{\hphm}}{\taueff{\Dzb}{\hphm} + \taueff{\Dz}{\hphm}}\,
  \simeq \left[ \onehalf (A_m+A_d) y \cos \phi - x \sin \phi \right],
\end{equation}
where $\tau_{\mathrm{eff}}$ is the effective lifetime, $\mathrm{h}$ can be either a pion or kaon, and
\begin{equation}
    A_m = \frac{\magsq{q/p} - \magsq{p/q}}{\magsq{q/p} + \magsq{p/q}}, \,
    A_d = \frac{\magsq{A_f/\bar{A}_f} - \magsq{\bar{A}_f/A_f}}{\magsq{A_f/\bar{A}_f} + \magsq{\bar{A}_f/A_f}},\,
    \phi = \mathrm{arg}\left(\frac{q}{p}\frac{\bar{A}_f}{A_f}\right),
\end{equation}
with $A_f$ ($\bar{A}_f$) the amplitude of the \Dz (\Dzb) meson decaying to the given final state.

Similarly to the analysis method discussed in Sec. \ref{sec:WSKpi} the decay chain \dstdpi is used to obtain the flavor of the \Dz candidate at production\cite{lhcb:AGamma2013}. The \KpKm and \pipi final states are both analyzed. For the \KpKm final state backgrounds from partially reconstructed three-body \Dz decays contribute in addition to combinatorics. These are distinguished by a fit to the distribution of \md and $\deltam \equiv \mdstar - \md$. 

\begin{figure}
  \centering
  \includegraphics[width = 0.4\textwidth]{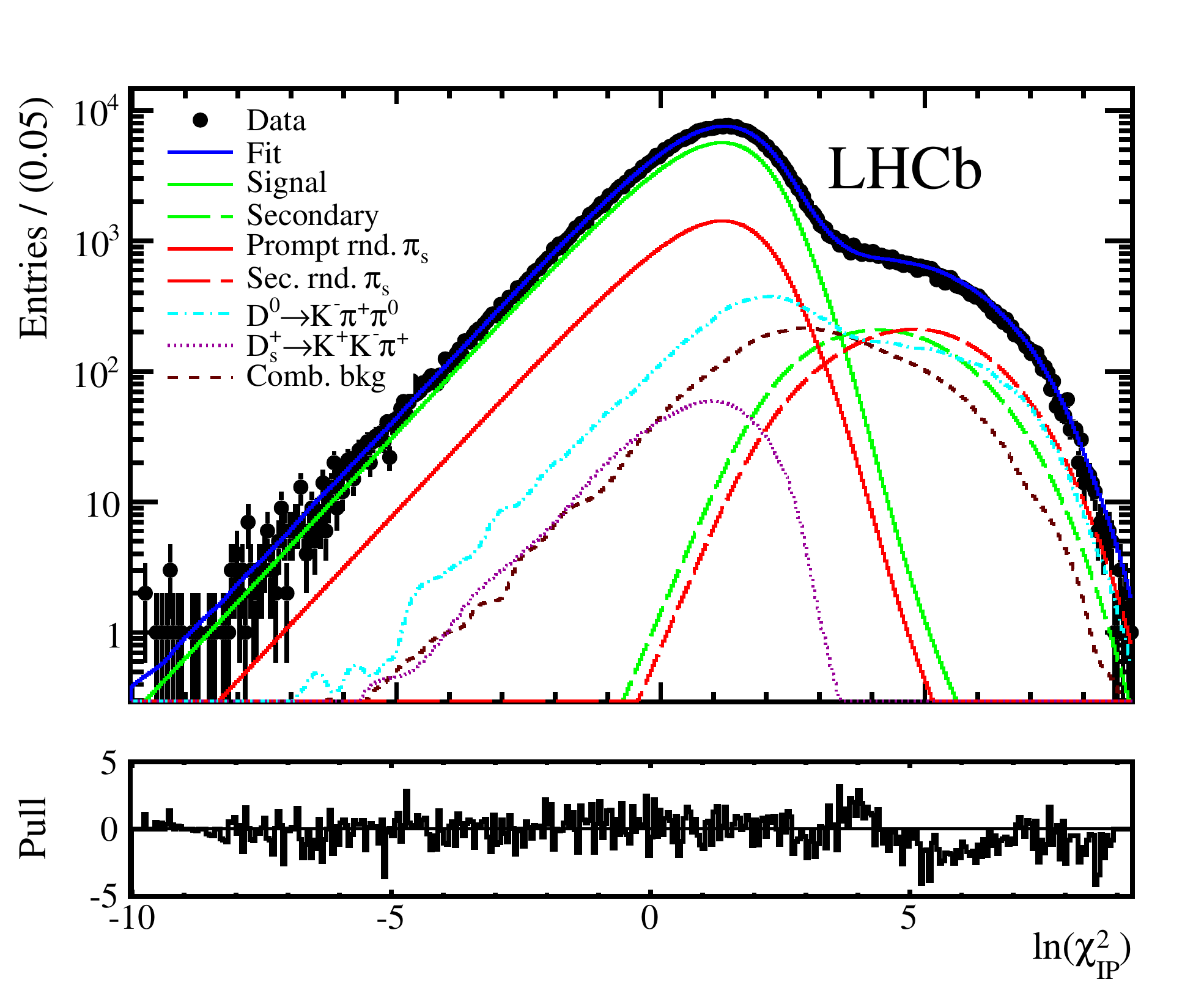}
  \includegraphics[width = 0.4\textwidth]{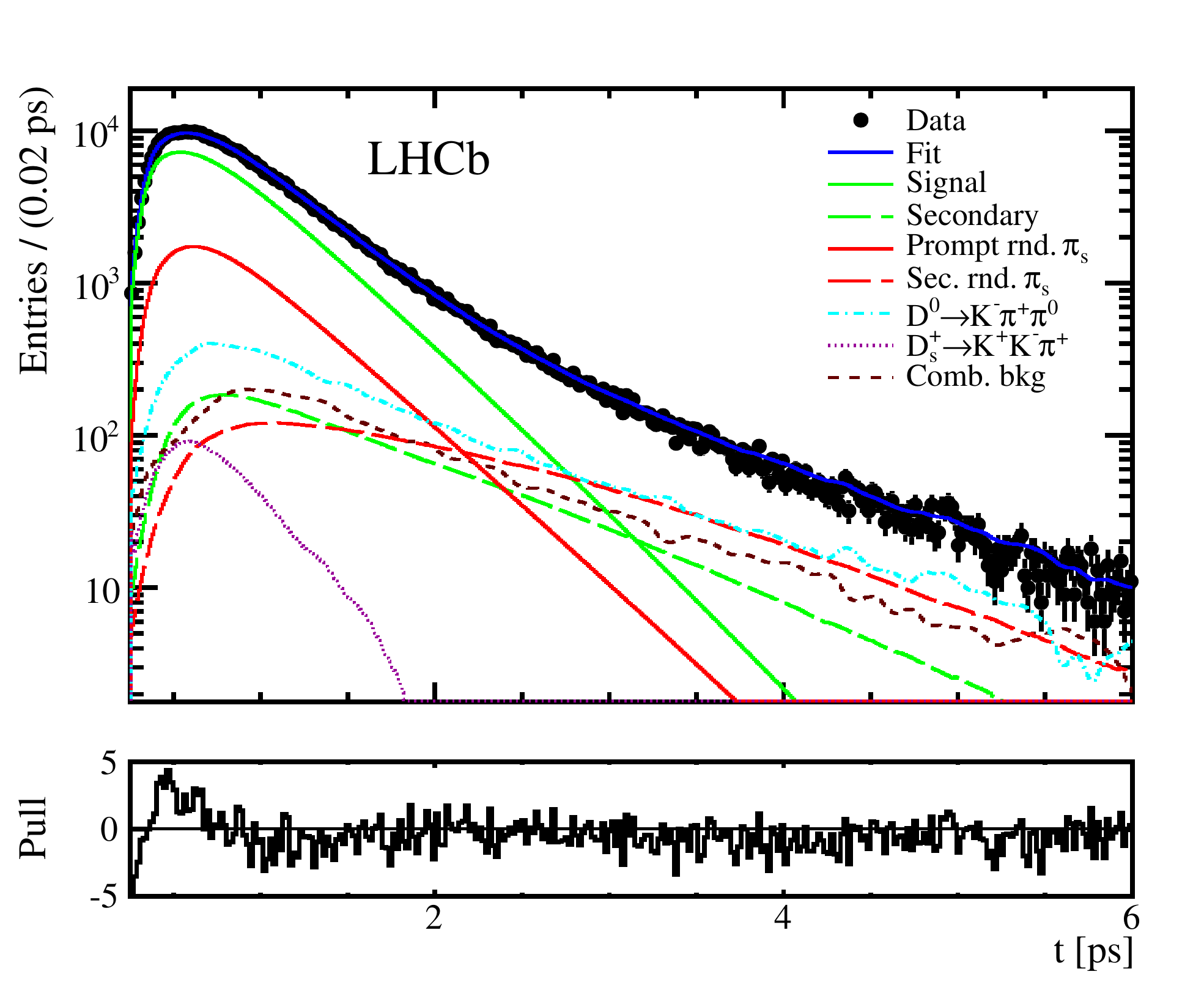}

  \caption[Fits to IP \chisq and decay time]{Examples of fits to the impact parameter \chisq and decay-time distributions for the \KpKm final state.}
  \label{fig:AGammaTimeFits}
\end{figure}

The effective lifetime is obtained by performing an unbinned maximum likelihood fit to the decay-time distribution using a data-driven, per-candidate method to correct for the biasing effect of the candidate selection\cite{lhcb_yCPAGamma}. Additionally, the distribution of the IP \chisq of the \Dz candidates is used to distinguish the background from \decay{\PB}{\Dz X} decays. Examples of fits to the \Dz IP \chisq and decay-time distributions are shown in Fig. \ref{fig:AGammaTimeFits}.

From the 2011 dataset these fits give
\begin{equation*}
    \agamma^{KK} = \xtene{(-0.35 \pm 0.62 \pm 0.12)}{-3},\,
    \agamma^{\pion\pion} = \xtene{(0.33 \pm 1.03 \pm 0.14)}{-3},
\end{equation*}
where dominant systematics arise from the accuracy of the acceptance correction and the modeling of the backgrounds. Thus, no evidence for indirect CP violation is found.


\section{Direct CP Violation in \decay{\Dz}{\hphm} Decays}
\label{sec:DACP}

The parameter
\begin{equation}
  \Delta \ACP = \ACP(\decay{\Dz}{\KpKm}) - \ACP(\decay{\Dz}{\pipi}),
\end{equation}
where \ACP is the time-integrated CP asymmetry of the given decay, gives direct access to direct CP violation as the production asymmetry of the \Dz meson cancels in the difference. To ensure full cancellation of production and detection asymmetries between the \KpKm and \pipi final states the \KpKm candidates are weighted so that their kinematic distributions match those of \pipi candidates.

Two independent datasets are used to measure $\Delta \ACP$: one in which the flavor of the \Dz candidate at production is determined using \dstdpi decays\cite{lhcb:DACP2013}, comprising predominantly prompt \Dz decays, and one in which the decay \decay{\Bm}{\Dz \mun} is reconstructed, with the charge of the \mun giving the flavor of the \Dz candidate\cite{lhcb:SLDACP2013}. The yields are determined by fits to the \deltam distribution for the pion tagged sample, and \md for the muon tagged sample.

Using the 2011 dataset the measurements obtained are
\begin{equation*}
  \Delta\ACP(\pion\,\mathrm{tagged}) = -0.34 \pm 0.15 \pm 0.10 \%,\, \Delta\ACP(\mmu\,\mathrm{tagged}) = +0.49 \pm 0.30 \pm 0.14 \%,
\end{equation*}
where the \pion tagged measurement is preliminary. This gives an average of 
\begin{equation*}
  \Delta\ACP = -0.15 \pm 0.16 \%.
\end{equation*}
Thus no evidence of direct CP violation is found.

\section{Rare Decays}
\label{sec:rare_decays}
The search for decays of \dtopimumu is sensitive to new physics as the process \decay{\c}{\u\mumu} is predicted to have a branching fraction of \otene{-9} in the SM\cite{Paul:c2dmumuBF}. Similarly, the lepton number violating decays \dtopimumuss are forbidden in the SM but can occur in some new physics scenarios, \eg with the existence of Majorana neutrinos. 

\begin{figure}
  \centering
  \includegraphics[height = 0.3\textwidth]{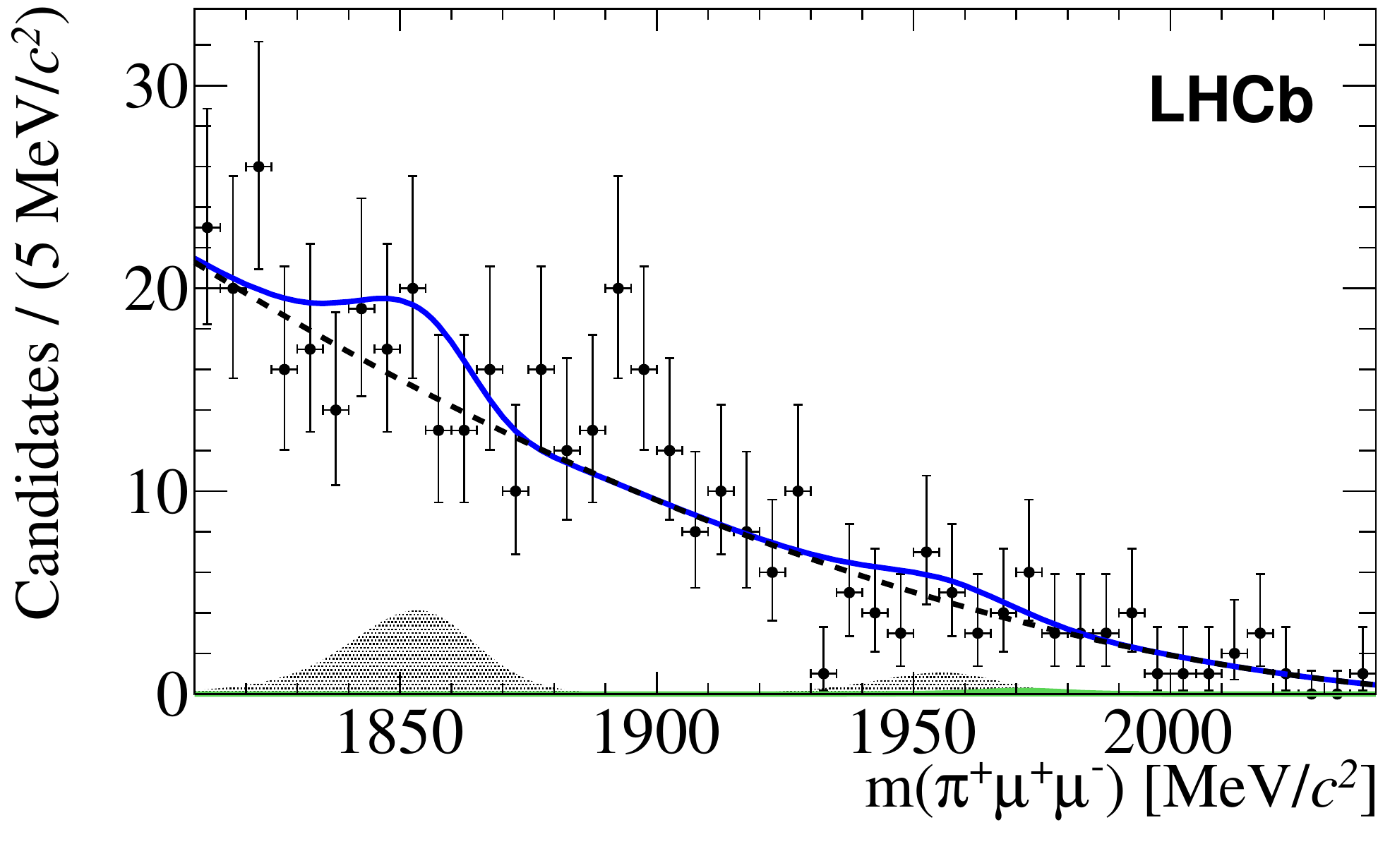}
  \includegraphics[height = 0.3\textwidth]{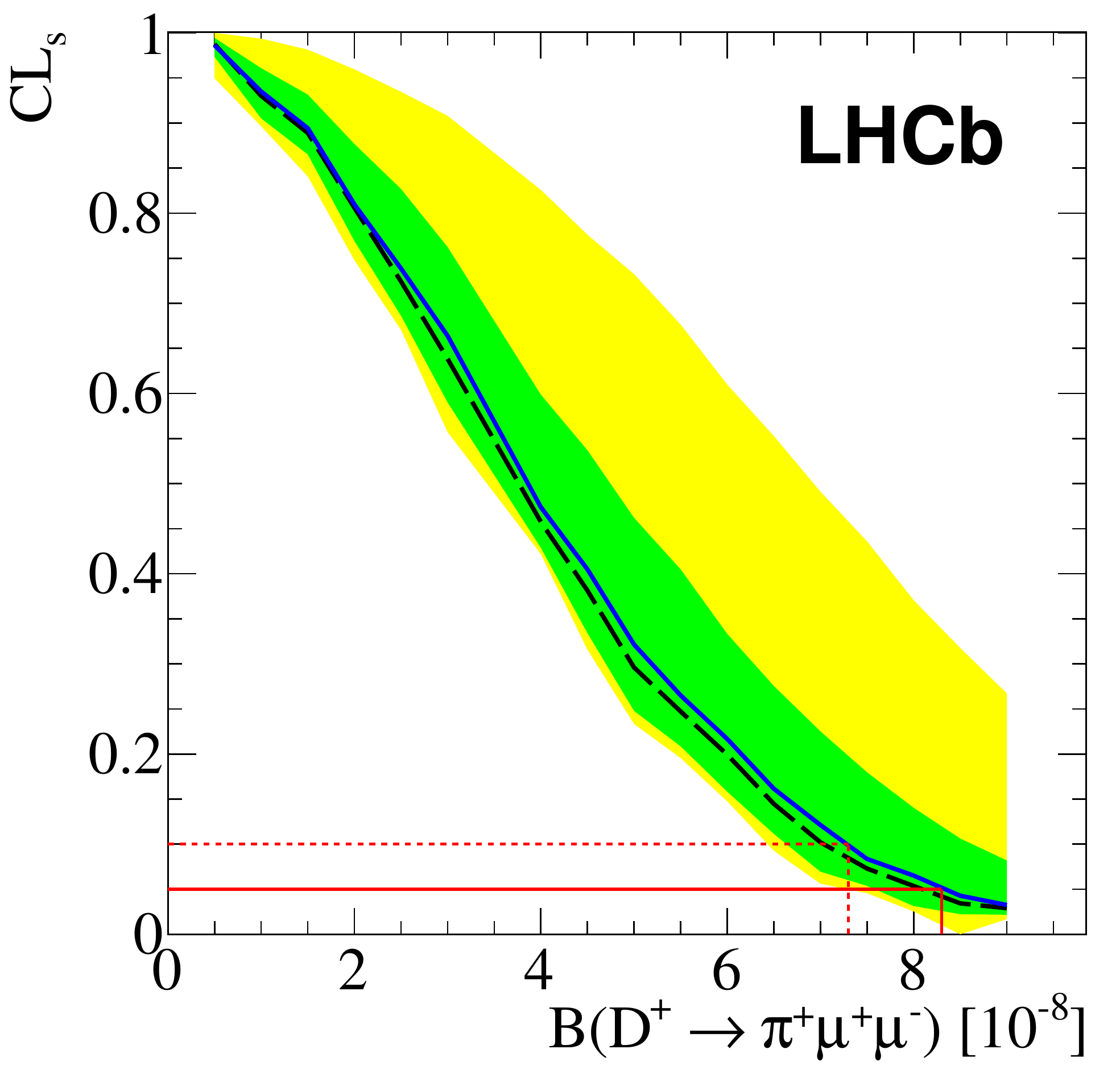}
  \caption[$\pion\mmu\mmu$ mass fit and confidence limits.]{(Left) the fit to the $\pip\mumu$ invariant mass distribution in the low \mumu mass region, with background from \decay{\Dp}{\pip\pipi} shown in solid gray; and (right) the confidence limits for $\BR(\decay{\Dp}{\pip\mumu})$.}
  \label{fig:pimumuplots}
\end{figure}

The distributions of $\pion\mmu\mmu$ invariant mass are fitted in bins of \mumu ($\pim\mup$) invariant mass in order to determine the yields\cite{lhcb:dtopimumu2013}. The limits on the signal yields are normalized to the yields found in the $m_{\mumu}$ region containing the \Pphi resonance and combined with the well known $\BR(\decay{\Dp_{(s)}}{\pip\decay{\Pphi(}{\mumu)}})$ to obtain limits on the partial branching fractions of \dtopimumuboth in each bin of $m_{\mumu}$ ($m_{\pim\mup}$). The total branching fractions of non-resonant decays are also constrained by combining information in bins of $m_{\mumu}$ ($m_{\pim\mup}$), assuming a phase-space model.
An example of a mass fit and branching fraction confidence limits for \decay{\Dp}{\pip\mumu} decays are shown in Fig. \ref{fig:pimumuplots}. 

Using the 2011 dataset the resulting 90 \% confidence limits on the total non-resonant branching fractions are
\begin{align*}
&\BR(\decay{\Dp}{\pip\mumu}) < \xtene{7.3}{-8},\, &\BR(\decay{\Ds}{\pip\mumu}) < \xtene{4.1}{-7}, \nonumber \\
&\BR(\decay{\Dp}{\pim\mup\mup}) < \xtene{2.2}{-8},\, &\BR(\decay{\Ds}{\pim\mup\mup}) < \xtene{1.2}{-7},
\end{align*}
which give a factor of roughly \tene{-2} improvement over the previous best limits. 

A similar search for \decay{\Dz}{\mumu} decays using the 2011 dataset finds \mbox{$\BR(\decay{\Dz}{\mumu}) < \xtene{6.2}{-9}$}
at 90 \% confidence\cite{lhcb:dtomumu2013}. Thus, no evidence for new physics has yet been found via rare decays.

\section{Conclusions}
\label{sec:conclusions}
Recent results on mixing, CP violation and rare decays in charm physics from the \lhcb experiment were presented. Study of \wskpi decays has provided the highest precision measurements of the mixing parameters \xpsq and \yp to date, and of CP violation in this decay mode. Direct and indirect CP violation in the \Dz system have been probed to a sensitivity of around \tene{-3} using \decay{\Dz}{\hphm} decays and found to be consistent with zero. Searches for the rare decays \dtopimumu, \dtopimumuss and \decay{\Dz}{\mumu} have found no evidence of signal, but have set the best limits on branching fractions to date. Thus, while no evidence for physics beyond the Standard Model has yet been found, the \lhcb experiment is firmly establishing itself as a world leader in high precision charm physics measurements.

\bibliographystyle{ws-ijmpcs}
\bibliography{\inputdir/bibliography.bib}

\end{document}

%% file: lhcb-symbols-preamble.tex
\usepackage{ifthen} 
\newboolean{uprightparticles}
\setboolean{uprightparticles}{true} 
\usepackage{xspace}     
\usepackage{amsfonts}
\usepackage{upgreek}

\newboolean{pdflatex}
\setboolean{pdflatex}{true} 

%% file: lhcb-symbols-def.tex
\def\lhcb {LHCb\xspace}
\def\ux85 {UX85\xspace}

\def\tt     {TT\xspace}

\ifthenelse{\boolean{uprightparticles}}%
{

 \def\Pmu         {\ensuremath{\upmu}\xspace}

 \def\Ppi         {\ensuremath{\uppi}\xspace}

 \def\Pphi        {\ensuremath{\upphi}\xspace}

 \def\PDelta      {\ensuremath{\Delta}\xspace}                 
 \def\PXi      {\ensuremath{\Xi}\xspace}                 
 \def\PLambda      {\ensuremath{\Lambda}\xspace}                 
 \def\PSigma      {\ensuremath{\Sigma}\xspace}                 
 \def\POmega      {\ensuremath{\Omega}\xspace}                 
 \def\PUpsilon      {\ensuremath{\Upsilon}\xspace}                 
 
 \def\PB      {\ensuremath{\mathrm{B}}\xspace}                 
                  
 \def\PD      {\ensuremath{\mathrm{D}}\xspace}

 \def\PK      {\ensuremath{\mathrm{K}}\xspace}

 \def\Pb      {\ensuremath{\mathrm{b}}\xspace}

 \def\Ph      {\ensuremath{\mathrm{h}}\xspace}                 
 \def\Pi      {\ensuremath{\mathrm{i}}\xspace}

 \def\Pu      {\ensuremath{\mathrm{u}}\xspace}

}
{

 \def\Pmu         {\ensuremath{\mu}\xspace}

 \def\Ppi         {\ensuremath{\pi}\xspace}

 \def\Pphi        {\ensuremath{\phi}\xspace}

 \mathchardef\PDelta="7101
 \mathchardef\PXi="7104
 \mathchardef\PLambda="7103
 \mathchardef\PSigma="7106
 \mathchardef\POmega="710A
 \mathchardef\PUpsilon="7107
                  
 \def\PB      {\ensuremath{B}\xspace}                 
                  
 \def\PD      {\ensuremath{D}\xspace}

 \def\PK      {\ensuremath{K}\xspace}

 \def\Pb      {\ensuremath{b}\xspace}

 \def\Ph      {\ensuremath{h}\xspace}                 
 \def\Pi      {\ensuremath{i}\xspace}

 \def\Pu      {\ensuremath{u}\xspace}

}

\def\mmu        {\ensuremath{\Pmu}\xspace}
\def\mup        {\ensuremath{\Pmu^+}\xspace}
\def\mun        {\ensuremath{\Pmu^-}\xspace} 
\def\mumu       {\ensuremath{\Pmu^+\Pmu^-}\xspace}

\def\u     {\ensuremath{\Pu}\xspace}

\def\c     {\ensuremath{\Pc}\xspace}
\def\cbar  {\ensuremath{\overline \c}\xspace}
\def\ccbar {\ensuremath{\c\cbar}\xspace}
\def\b     {\ensuremath{\Pb}\xspace}

\def\pion  {\ensuremath{\Ppi}\xspace}

\def\pip   {\ensuremath{\pion^+}\xspace}
\def\pim   {\ensuremath{\pion^-}\xspace}
\def\pipi  {\ensuremath{\pion^+\pion^-}\xspace}
\def\pipm  {\ensuremath{\pion^\pm}\xspace}
\def\pimp  {\ensuremath{\pion^\mp}\xspace}

\def\kaon  {\ensuremath{\PK}\xspace}
  \def\Kbar  {\kern 0.2em\overline{\kern -0.2em \PK}{}\xspace}

\def\Kz    {\ensuremath{\kaon^0}\xspace}
\def\Kzb   {\ensuremath{\Kbar^0}\xspace}
\def\KzKzb {\ensuremath{\Kz \kern -0.16em \Kzb}\xspace}
\def\Kp    {\ensuremath{\kaon^+}\xspace}
\def\Km    {\ensuremath{\kaon^-}\xspace}
\def\Kpm   {\ensuremath{\kaon^\pm}\xspace}

\def\KpKm  {\ensuremath{\Kp \kern -0.16em \Km}\xspace}

  \def\Dbar    {\kern 0.2em\overline{\kern -0.2em \PD}{}\xspace}
\def\D       {\ensuremath{\PD}\xspace}

\def\Dz      {\ensuremath{\D^0}\xspace}
\def\Dzb     {\ensuremath{\Dbar^0}\xspace}
\def\DzDzb   {\ensuremath{\Dz {\kern -0.16em \Dzb}}\xspace}
\def\Dp      {\ensuremath{\D^+}\xspace}
\def\Dm      {\ensuremath{\D^-}\xspace}

\def\DpDm    {\ensuremath{\Dp {\kern -0.16em \Dm}}\xspace}

\def\Dstarp  {\ensuremath{\D^{*+}}\xspace}

\def\Ds      {\ensuremath{\D^+_s}\xspace}

\def\B       {\ensuremath{\PB}\xspace}
  \def\Bbar    {\kern 0.18em\overline{\kern -0.18em \PB}{}\xspace}

\def\Bub     {\ensuremath{\B^-}\xspace}

\def\Bm      {\ensuremath{\Bub}\xspace}

  \def\Y#1S{\ensuremath{\PUpsilon{(#1S)}}\xspace}

\def\BR         {{\ensuremath{\cal B}\xspace}}

\newcommand{\decay}[2]{\ensuremath{#1\!\to #2}\xspace}         

\def\to                 {\ensuremath{\rightarrow}\xspace}

\def\order   {\ensuremath{\mathcal{O}}\xspace}

\newcommand{\ACP}{\ensuremath{{\cal A}^{\rm CP}}\xspace}

\def\AT#1     {\ensuremath{A_T^{#1}}\xspace}           

\def\C#1      {\ensuremath{\mathcal{C}_{#1}}\xspace}                       
\def\Cp#1     {\ensuremath{\mathcal{C}_{#1}^{'}}\xspace}                    
\def\Ceff#1   {\ensuremath{\mathcal{C}_{#1}^{\mathrm{(eff)}}}\xspace}        
\def\Cpeff#1  {\ensuremath{\mathcal{C}_{#1}^{'\mathrm{(eff)}}}\xspace}       
\def\Ope#1    {\ensuremath{\mathcal{O}_{#1}}\xspace}                       
\def\Opep#1   {\ensuremath{\mathcal{O}_{#1}^{'}}\xspace}                    

\def\agamma     {\ensuremath{A_{\Gamma}}\xspace}

\newcommand{\ket}[1]{\ensuremath{|#1\rangle}}              

\newcommand{\tev}{\ensuremath{\mathrm{\,Te\kern -0.1em V}}\xspace}
\newcommand{\gev}{\ensuremath{\mathrm{\,Ge\kern -0.1em V}}\xspace}
\newcommand{\mev}{\ensuremath{\mathrm{\,Me\kern -0.1em V}}\xspace}
\newcommand{\kev}{\ensuremath{\mathrm{\,ke\kern -0.1em V}}\xspace}
\newcommand{\ev}{\ensuremath{\mathrm{\,e\kern -0.1em V}}\xspace}
\newcommand{\gevc}{\ensuremath{{\mathrm{\,Ge\kern -0.1em V\!/}c}}\xspace}
\newcommand{\mevc}{\ensuremath{{\mathrm{\,Me\kern -0.1em V\!/}c}}\xspace}
\newcommand{\gevcc}{\ensuremath{{\mathrm{\,Ge\kern -0.1em V\!/}c^2}}\xspace}
\newcommand{\gevgevcccc}{\ensuremath{{\mathrm{\,Ge\kern -0.1em V^2\!/}c^4}}\xspace}
\newcommand{\mevcc}{\ensuremath{{\mathrm{\,Me\kern -0.1em V\!/}c^2}}\xspace}

\def\invfb   {\ensuremath{\mbox{\,fb}^{-1}}\xspace}

\def\fs   {\ensuremath{\rm \,fs}\xspace}

\def\order{{\ensuremath{\cal O}}\xspace}
\newcommand{\chisq}{\ensuremath{\chi^2}\xspace}

\def\gsim{{~\raise.15em\hbox{$>$}\kern-.85em
          \lower.35em\hbox{$\sim$}~}\xspace}
\def\lsim{{~\raise.15em\hbox{$<$}\kern-.85em
          \lower.35em\hbox{$\sim$}~}\xspace}

\def\sqs   {\ensuremath{\protect\sqrt{s}}\xspace}

\def\ptot       {\mbox{$p$}\xspace}

\def\tell1  {TELL1\xspace}
\def\ukl1   {UKL1\xspace}

\newcommand{\eg}{\mbox{\itshape e.g.}\xspace}

%% file: analysis-symbols-def.tex
\def\deltam     {\ensuremath{\Delta m}\xspace}

\def\hphm       {\ensuremath{\Ph^+ \Ph^-}\xspace}

\newcommand{\taueff}[2]{\ensuremath{\tau_{\mathrm{eff}}( \decay{#1}{#2} )}\xspace}

\def\c          {\ensuremath{c}\xspace}

\newcommand{\magnitude}[1]{\ensuremath{\left|{#1}\right|}\xspace}
\newcommand{\magsq}[1]{\ensuremath{\magnitude{#1}^2}\xspace}

\newcommand{\orderof}[1]{\ensuremath{\order(#1)}\xspace}

\newcommand{\tene}[1]{\ensuremath{10^{#1}}\xspace}
\newcommand{\xtene}[2]{\ensuremath{#1 \times \tene{#2}}\xspace}
\newcommand{\otene}[1]{\orderof{\tene{#1}}}

\def\onehalf		{\ensuremath{\frac{1}{2}}\xspace}

\def\C	{\ensuremath{C}\xspace}

\newcommand{\sqrtseq}[1]{\ensuremath{\sqs = #1}\xspace}

%% file: general-latex.tex
\newcommand{\dquote}[1]{``#1''}